\documentclass[a4paper]{article}

\usepackage[a4paper, total={155mm,260mm}]{geometry}

\usepackage{amsmath,amssymb,amsfonts}
\usepackage{algorithmic}
\usepackage{graphicx}
\usepackage{textcomp}

\usepackage{xcolor}
\usepackage{csvsimple}
\usepackage{subfig}
\usepackage{enumitem}
\usepackage{url}
\usepackage{wrapfig}
\usepackage{dirtytalk}
\usepackage{multirow}
\usepackage{tikz}

\usepackage{listings}
\usepackage{authblk}
\providecommand{\keywords}[1]
{
  \small	
  \textbf{\textit{Keywords---}} #1
}

\title{Nori: Concealing the Concealed Identifier in 5G}

\author[]{John Preuß Mattsson}
\author[]{Prajwol Kumar Nakarmi}
\affil[]{Ericsson, Sweden}
\affil[]{\textit {\{john.mattsson, prajwol.kumar.nakarmi\}@ericsson.com}}

\date{June, 2021}

\begin{document}

\maketitle

\begin{tikzpicture}[overlay]
  \node[draw, fill=white, thick, text width=17cm,minimum height=1cm] (b) at (7,7){\footnotesize The paper is to be published in the 16th International Conference on Availability, Reliability and Security (ARES 2021)\\August 17–20, 2021, Vienna, Austria. Please cite that version. Publication rights licensed to ACM.\\ACM ISBN 978-1-4503-9051-4/21/08. https://doi.org/10.1145/3465481.3470076. };
\end{tikzpicture}

\begin{abstract}
	IMSI catchers have been a long standing and serious privacy problem in pre-5G mobile networks. To tackle this, 3GPP introduced the Subscription Concealed Identifier (SUCI) and other countermeasures in 5G. In this paper, we analyze the new SUCI mechanism and discover that it provides very poor anonymity when used with the variable length Network Specific Identifiers (NSI), which are part of the 5G standard. When applied to real-world name length data, we see that SUCI only provides 1-anonymity, meaning that individual subscribers can easily be identified and tracked. We strongly recommend 3GPP and GSMA to standardize and recommend the use of a padding mechanism for SUCI before variable length identifiers get more commonly used. We further show that the padding schemes, commonly used for network traffic, are not optimal for padding of identifiers based on real names. We propose a new improved padding scheme that achieves much less message expansion for a given $k$-anonymity.
\end{abstract}

\section{Introduction}
Cellular devices such as mobile phones, tablets, and wearables have become more pervasive. Their role in the leakage of Personally Identifiable Information (PII) is also scrutinized more than ever, and rightfully so. One main category of PII consists of the permanent International Mobile Subscriber Identity (IMSI). While all pre-5G cellular networks (2G, 3G, and 4G) assigned and used temporary identifiers, the permanent IMSI was sometimes sent in cleartext over the radio interface. Active attacker could also trick cellular devices to send their IMSIs over the radio interface using so-called \say{IMSI catchers}, and identify as well as track victims \cite{3gpp_33899, Shaik2016PracticalAA, Borgaonkar2018NewPT, Hussain20195GReasonerAP, Hussain2019PrivacyAT}. In order to solve this problem, 5G introduced the Subscription Concealed Identifier (SUCI) mechanism used to encrypt Subscription Permanent Identifier (SUPI) as well as other protection mechanisms like strict refreshment of temporary identifiers, decoupling of the permanent identifier from the paging mechanism, and secure radio redirections \cite{eri_blog_battle}. SUCI is calculated by using  Elliptic Curve Integrated Encryption Scheme (ECIES) \cite{ecies1, ecies2}.

In this paper we discover a vulnerability (Section \ref{sec:suci-anal}) in how 3GPP has specified the ECIES profiles for SUCI in 5G security standard TS 33.501 \cite{3gpp_33501}. As SUCI encryption use AES in counter mode (CTR), the ciphertext is of same length as the input plaintext. This means means that SUCIs -- even though fresh each time -- could be linked to each other or with SUPIs based on their lengths. Therefore, SUCI does not provide indistinguishability for variable length identifiers since it leaks PII that can be practically useful for an attacker. 

We propose padding as a fix to the above-mentioned problem. This means that SUPI is padded before encryption and the exposure of PII from SUCI is alleviated. In order to assess the effect of padding, we collected real-world name length data and used information theory metrics to empirically quantify message expansion and privacy protection. We present analysis on two sets of anonymized real-world name length data from (a) Swedish population, and (b) a multi-national Company's employees in Sweden, China, India, and the USA. These (anonymized) datasets are representative of 5G use cases and are described in Section \ref{sec:real_data}. We investigated five padding schemes inspired from \cite{ndss, rfc8467} as well as a new \say{tail-aware block-length} padding scheme that we designed (Section \ref{sec:analysis}). For many types of distributions, the \say{tail-aware block-length} padding scheme is much more efficient than previous padding schemes as it in many cases provides better anonymity with less message expansion.

We call our work \textbf{Nori}\footnote{Because 3GPP decided that SUCI is pronounced as SU-SHI.}. We have presented Nori to 5G Security Task Force of GSMA’s Fraud and Security Group (FASG) \cite{gsma_security}. GSMA FASG acknowledged that the vulnerability is real and should be fixed. We received positive gestures that this work will likely be pursued in 3GPP, aiming for Release-18. Also, based on the work in this paper, IETF HPKE \cite{irtf-cfrg-hpke-08} (proposed to be used in TLS 1.3 \cite{rfc8446} and MLS \cite{ietf-mls-protocol-11}) has introduced a recommendation on padding.  

In summary, our contributions are:
\begin{itemize}    
    \item Discover an information leakage in the SUCI construction with variable-length identifiers and propose padding as a fix.
    \item Propose a new tail-aware block-length padding scheme that minimizes message expansion.
    \item Provide real-world data on name lengths for all people in Sweden as well as a multi-national company.    
    \item Empirical evaluation of different padding schemes on real-world name length data.  
    \item Standardization impact: IETF HPKE has introduced recommendation on padding; 3GPP will likely introduce padding for SUCIs in Release-18.
\end{itemize}

\section{Background}
5G is the fifth generation of mobile networks standardized by 3GPP \cite{3gpp_23501, 3gpp_23502, 3gpp_38401}. It is an evolution of earlier generations called 4G/LTE, 3G/UMTS, and 2G/GSM/GPRS. Each subscription in a 5G network is identified by a unique long-term identifier called Subscription Permanent Identifier (SUPI) \cite{3gpp_23003}. For privacy protection of SUPI, 3GPP introduced Subscription Concealed Identifier (SUCI), specified in 3GPP TS 33.501 \cite{3gpp_33501} and TS 23.003 \cite{3gpp_23003}. 

\begin{figure*}[h]
      \centering
       \includegraphics[width=13cm]{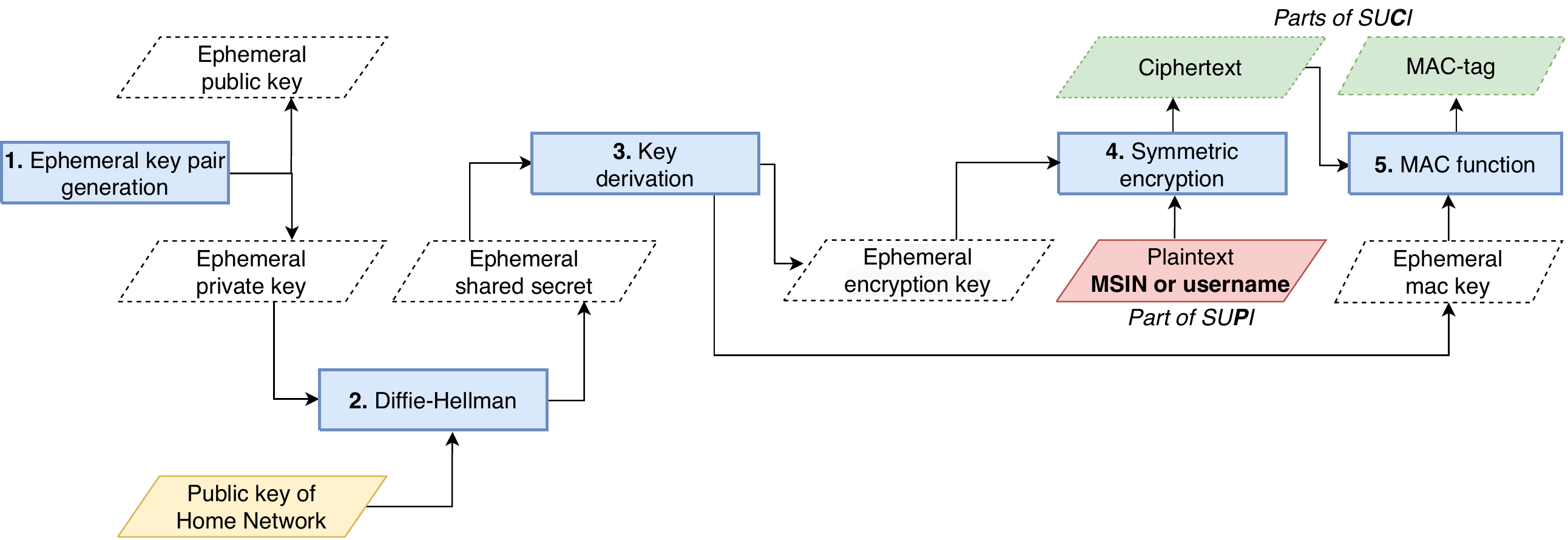}
       \caption{SUCI calculation using ECIES.
      \label{fig:5g_suci_ecies}}
\end{figure*}

SUCI encrypts SUPI with the Elliptic Curve Integrated Encryption Scheme (ECIES) scheme \cite{ecies1, ecies2}, as shown in Fig. \ref{fig:5g_suci_ecies}. ECIES is a hybrid scheme in which key exchange is based on asymmetric cryptography and key derivation and encryption are based on symmetric cryptography. ECIES is a probabilistic encryption scheme where the same plaintext encrypted multiple times produces completely different ciphertexts that cannot be linked to each other or the plaintext. 3GPP has standardized three protection schemes \cite{3gpp_33501}: Null-scheme, Profile A, and Profile B. The \say{null-scheme} does not do any actual encryption, rather produces the same output as the input. The Profile A and B use Curve25519 or secp256r1 together with AES-128-CTR and HMAC-SHA-256. 

SUCI is composed as follows where the SUPI type is either IMSI or Network Specific Identifier (NSI). NSI is in the format of Network Access Identifier (NAI) as defined in IETF RFC 7542 \cite{rfc7542}. Only the concealed identifier part is encrypted by SUCI. 
\[\texttt{SUCI = SUPI\;type\;||\;Home\;network\;identifier ||\;other\;parameters ||\;Concealed\;identifier}\]
When the SUPI is of type \textbf{IMSI}, the Home network identifier is composed of a 3-digit Mobile Country Code (MCC), and a 2--3-digit Mobile Network Code (MNC) and the concealed identifier contains encrypted 9--10-digit Mobile Subscription Identification Number (MSIN). When the SUPI is of type \textbf{NSI}, the Home network identifier is composed of a variable length string called the realm, and the concealed identifier contains a variable length encrypted string called the username.

\section{Real-World Name Length Data \label{sec:real_data}}
How NSI type SUPIs will be created in future 5G networks is yet to be seen. But it is likely that many networks will have them created from real-world names because earlier and current uses of such identifiers, e.g., in ISIMs (IP Multimedia Services Identity Module), have been based on real-world names. In this section, we present two name length data that we believe will be useful for other researchers in the field of privacy. 

\subsection{Name Length Data in Sweden}
With the kind help from Swedish government agency SCB (Swedish: Statistiska centralbyrån) \cite{SCB}, we got access to the name length data for the whole of Sweden as well as Stockholm Municipality as of 31 December 2019 (Appendix \ref{sec:scb_data}). We gathered two data sets, one with only the first and last names, and another with an additional (optional) maiden name. The name lengths are counted straight off with no spaces between different parts of the name. Their distributions, shown in Fig. \ref{fig:scb_dist_fit}, are very similar, therefore, we have only analyzed the data without the maiden name in the rest of the paper. With a population of 10 million people, it is a reasonable data set to analyze the subscriber privacy of a medium-sized 5G network operator.

We note that Swedish names can contain three non-ASCII characters `å', `ä', and `ö'. These non-ASCII characters are in our analysis assumed to have a one octet encoding as they are typically replaced by `a' or `o' when used in NAIs and email addresses. A two octet UTF-8 encoding of `å', `ä', and `ö' would slightly change the distributions.

\subsection{Name Length Data in a Company}
\begin{figure*}[]
  \centering
  \subfloat[Sweden and Stockholm \label{fig:scb_dist_fit}]{
    \includegraphics[height=5cm]{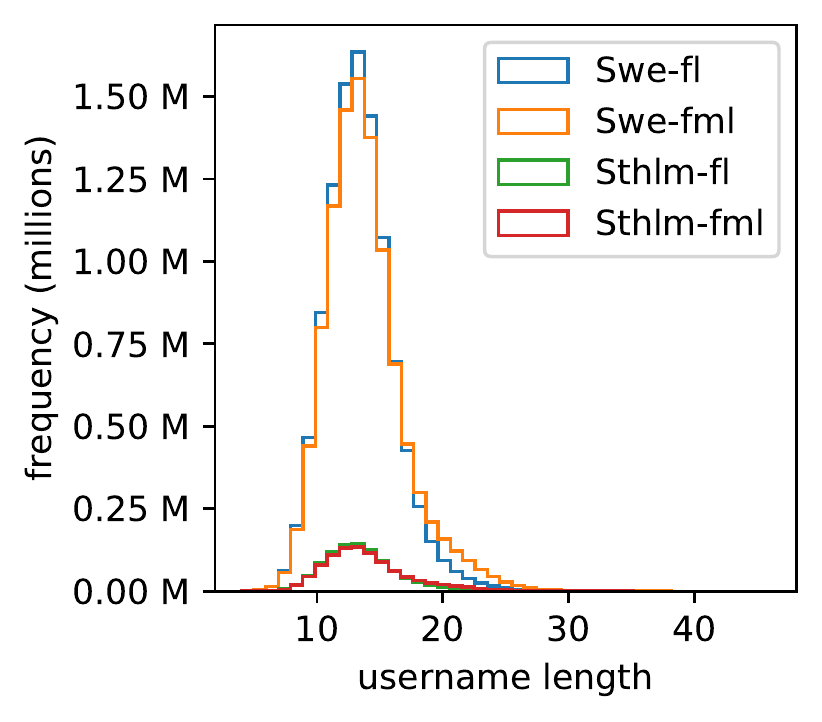}
  }
  \subfloat[A Company \label{fig:eri_dist_fit}]{
    \includegraphics[height=5cm]{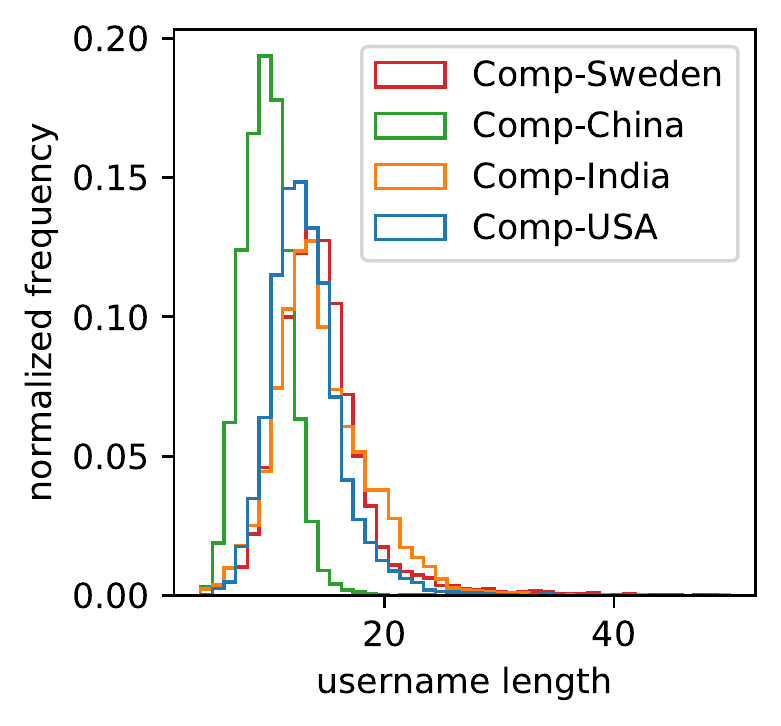}
  }
  \caption{Histogram for name lengths. fl = first name $||$ last name; fml = first name $||$ maiden name $||$ last name.}
\end{figure*}

5G is to a large degree designed for use in various industries and factories. It is expected that many industries using 5G will operate their own database with subscribers and use their own realm. To that end, we gathered anonymized name length data based on email addresses of a multi-national Company. We chose four of the countries -- Sweden, China, India, and the USA -- it is present in, that represent different languages and cultures. Their normalized frequency distributions are shown in Fig. \ref{fig:eri_dist_fit}. It can be seen that the distributions for Sweden, India, and the USA are fairly similar, while the distribution for China has much smaller mean and variance. With a few thousand employees in each country, this is a reasonable data set to analyze the subscriber privacy of a medium-sized industry using 5G.

We note that the usernames in the company email addresses had already been transliterated to ASCII characters, which is a very common practice. A UTF-8 encoding with non-ASCII character might significantly change the distribution for countries with non-roman characters, for which the recommendations in this paper might need to be adjusted. 

\section{Vulnerability in SUCI when applied to variable length SUPI \label{sec:suci-anal}}
By using ECIES, SUCI achieves the notion of \say{indistinguishability} or \say{semantic security}, meaning even a very capable attacker cannot distinguish the encrypted ciphertext from a random string. However, this indistinguishability game assumes fixed length plaintexts and if the assumption is not true, then the indistinguishability is broken. 

The vast majority of current 5G networks use the IMSI type SUPI where the MSIN has a fixed length for a given MCC. In that case, SUCI based on IMSI is also fixed length and provides indistinguishability. But when the NSI type SUPI is used, the username has a variable length. Therefore, SUCI based on NSI also has a variable length and indistinguishability no longer holds. An attacker gets perfect information regarding the length of the username.

Given the assumption that NSI type SUPIs will be created from real-world names, SUCI applied to the datasets in Section \ref{sec:real_data} shows poor anonymity for users with very short or very long names. In terms of $k$-anonymity \cite{privacy_metric}, even when applied to the very large datasets from the whole of Sweden and Stockholm, the SUCI only achieves 3-anonymity. The anonymity is likely even worse as the numbers for longest length in these datasets are known to exclude data that are potentially incorrect during petition and a correct dataset would likely only provide 1-anonymity. 1-anonymity means that an attacker can trivially identify or track at least one of the users. Similarly, SUCI applied to the company datasets from Sweden, USA, India, and China provide the worst possible 1-anonymity. 

The conclusion from this is that SUCI applied to NSI type identifiers created from real-world names provides very poor anonymity for users with unusual identifier lengths, i.e., very short or very long names.

\section{Applying Padding to Real-World Name Data \label{sec:analysis}}

To conceal the username length leaked by SUCI and make it harder for an attacker to distinguish SUCIs based on their lengths, we suggest padding the identifier before encryption. With reference to Fig. \ref{fig:5g_suci_ecies}, the plaintext has to be padded before Step 4.

\subsection{Padding Schemes}
We evaluated six padding schemes in total. Five of them were inspired from \cite{rfc8467, ndss}: (a) Block-length (\emph{blk-sz-min}), (b) Power-length (\emph{pwr-b-min}), (c) Random block-length (\emph{rndBlk-sz-blks-min}), (d) Random-length padding (\emph{rndLen-len}), and (e) Maximum-length (\emph{max-len}). 

\begin{figure}[h]
  \centering
  \includegraphics[height=2.5cm]{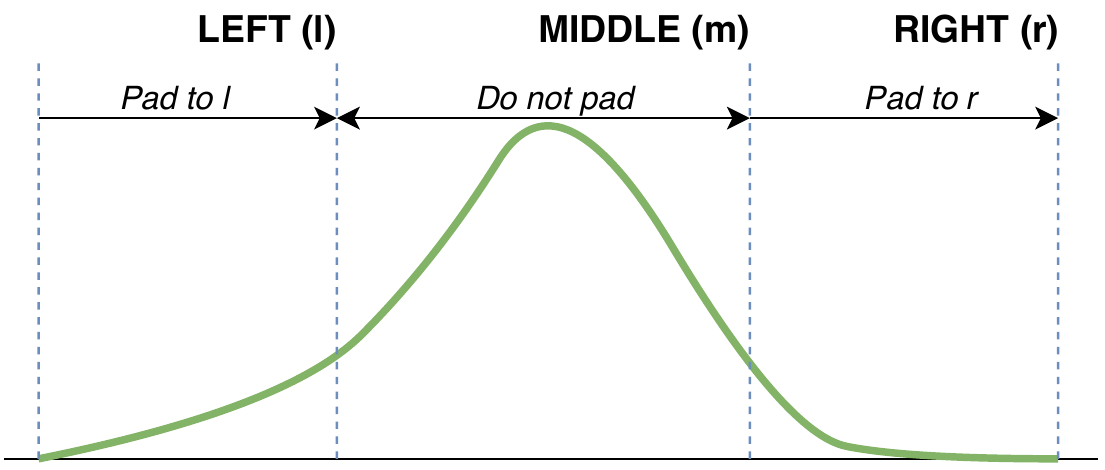}
  \caption{Illustration of the \emph{taBlk-l-m-r} padding. \label{fig:tablk}}
\end{figure}

The sixth padding scheme \say{tail-aware block-length} padding (\emph{taBlk-l-m-r}) is designed by us. The intuition behind it is that the tails of typical distributions have the lowest frequency (meaning lower anonymity), and which benefit from padding the most. The middle parts of distributions typically have much higher frequencies and padding those only contribute to message expansion without significant increase in privacy.  Therefore, we propose \emph{taBlk-l-m-r} padding done as shown in Fig. \ref{fig:tablk}, i.e., lengths below LEFT ($l$) are padded to $l$; lengths between $l$ and MIDDLE ($m$) are not padded; and lengths above $m$ are padded to RIGHT ($r$). By doing such selective padding, the overall message expansion is significantly reduced compared to other padding schemes that pad on all ranges of lengths.

\subsection{Evaluation of Padding Schemes \label{sec:eval_pad_scheme}}
We empirically evaluated the six padding schemes on the real-world name length data described in Section \ref{sec:real_data}. In total, we tested 868 padding instances with varying parameters (184 instances of \emph{blk}, 44 of pwr, 184 of \emph{rndBlk}, 32 of \emph{rndLen}, 420 of \emph{taBlk}, and 4 of \emph{maxL}). For \emph{maxL}, there is only one instance per dataset and everything is padded to the same size as the longest length in the dataset.

We assigned two attacker costs $\alpha_1$, $\alpha_2$, and a defender cost $\beta$ to each padding instance. The aim of the evaluation is to identify padding schemes that maximizes $\alpha_1$ and $\alpha_2$ while keeping $\beta$ low. The costs are defined below:

\begin{itemize}
  \item \textbf{Attacker Cost 1 ($\alpha_1$)}: Conditional entropy $H(U|P)$ \cite{privacy_metric}, where $U$ is the distribution before padding (unpadded) and $P$ is the distribution after padding. $\alpha_1$ represents the uncertainty about the outcome of $U$ when the outcome of $P$ is known, or as the expected number of bits needed to describe $U$ given that $P$ is known.

  \item \textbf{Attacker Cost 2 ($\alpha_2$)}: $k$-anonymity \cite{privacy_metric}, meaning the information for each person contained in a data set cannot be distinguished from at least $k-1$ people whose information also appear in the data set. $k$-anonymity is sometimes referred to as a \say{hiding in the crowd} guarantee. 

  \item \textbf{Defender Cost ($\beta$)}: Increased bandwidth as defined in \cite{ndss}, i.e., a weighted sum of all the padded lengths normalized by the unpadded lengths. In this paper, we only consider the length of identifiers while \cite{ndss} looks at the whole packets transported on the wire. If the identifiers are transported in larger packets, the size of padded identifiers might be small compared to the packet size.
\end{itemize}

Fig. \ref{fig:defendervsattacker} and \ref{fig:defendervsattacker2} show the $\alpha_1$, $\alpha_2$ vs $\beta$ plots for each padding instance and dataset. Each point in the plots represents a particular padding instance. Fig. \ref{fig:zoom} zooms into the plots for Sweden where $\beta$ is limited to 2.0, i.e., maximum message size expansion of double.

\subsection{Best Performing Padding Scheme}
\begin{figure*}[h!]
\centering  
\subfloat[$\alpha_1$ vs $\beta$ zoom-in for Sweden. \label{fig:defendervsattacker_zoom}]{
  \includegraphics[height=5cm]{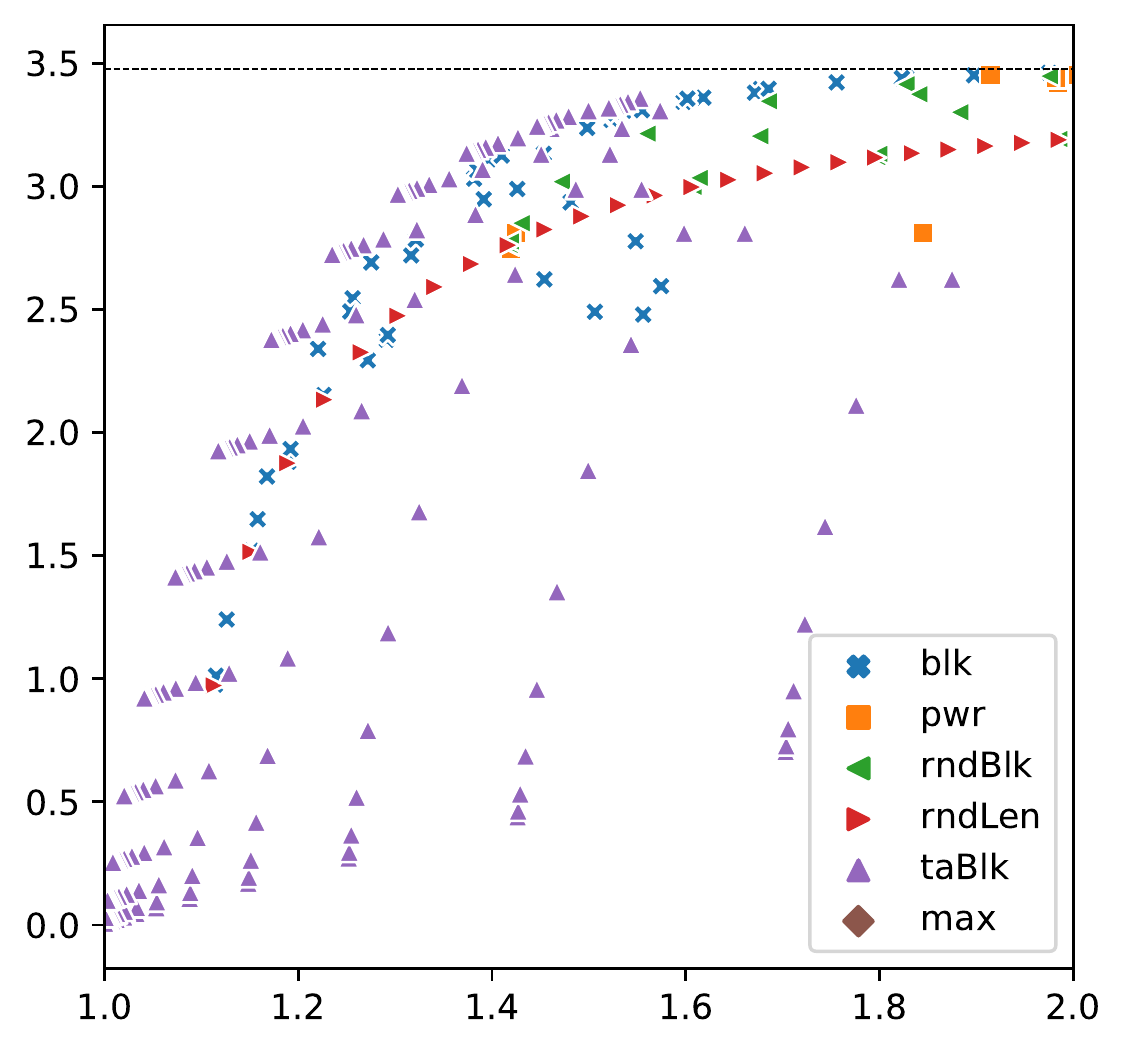}
}
\subfloat[Log of $\alpha_2$ vs $\beta$ zoom-in for Sweden. \label{fig:defendervsattacker2_zoom}]{
  \includegraphics[height=5cm]{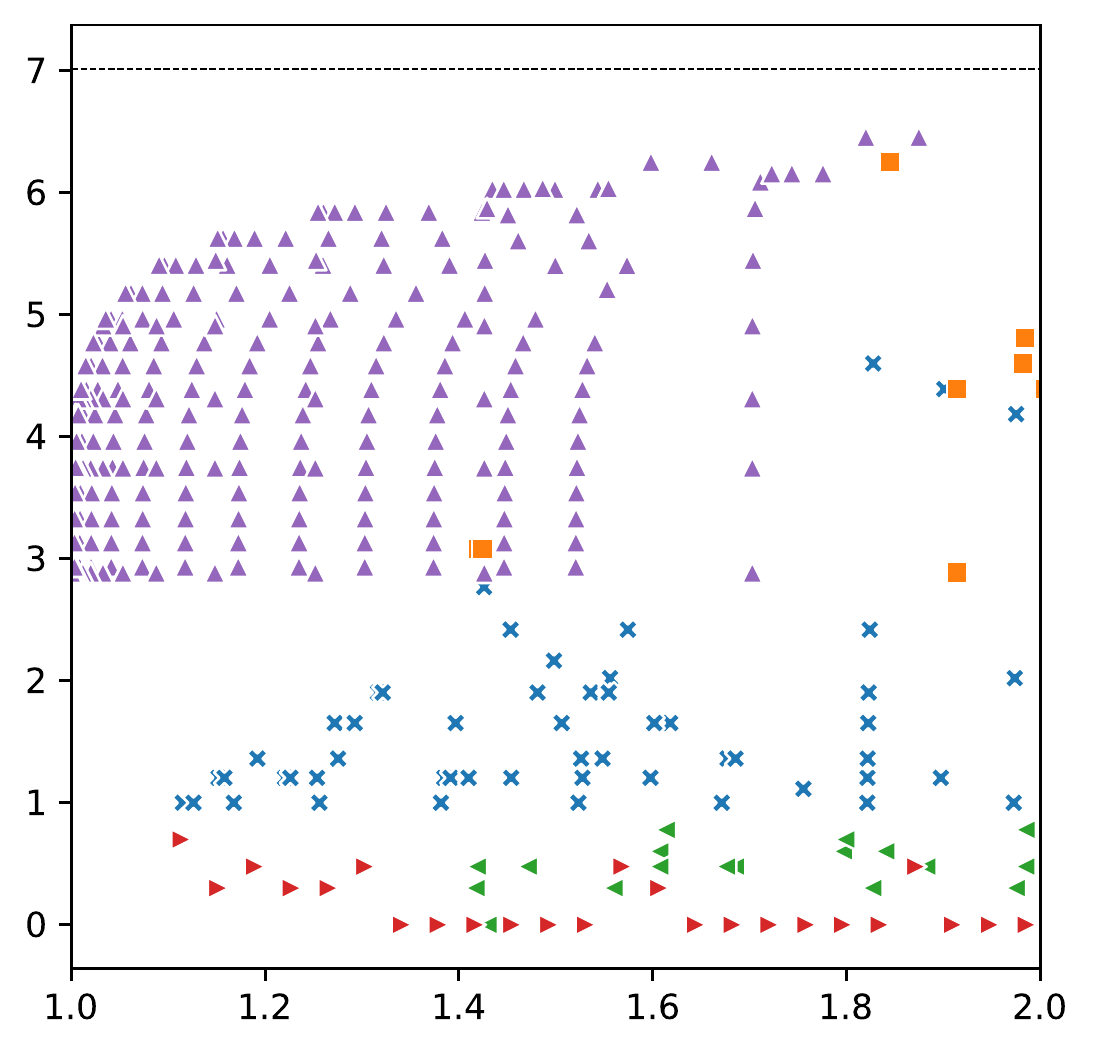}
}   
\caption{Zoom in plots for Sweden for $1\leq\beta\leq2$. Plot for \emph{max} is not present in this range of $\beta$. \label{fig:zoom}}  
\end{figure*}

Just looking at Fig. \ref{fig:zoom} it clear that for the intervals shown in the figure, \emph{taBlk-l-m-r} is performing much better than the other five padding schemes. While is is hard to put exact values on how much privacy protection ($\alpha_1$, $\alpha_2$) is needed and how much bandwidth ($\beta$) is acceptable. We believe that Fig. \ref{fig:zoom} covers the values of $\alpha_1$, $\alpha_2$, and $\beta$ for most practical deployments. In order to choose the best padding scheme and parameters that maximize $\alpha_1$ and $\alpha_2$ while minimizing $\beta$, we use two additional metrics as defined below:

\begin{figure*}[h!]
  \centering  
  \subfloat[Sweden]{
    \includegraphics[height=2.8cm]{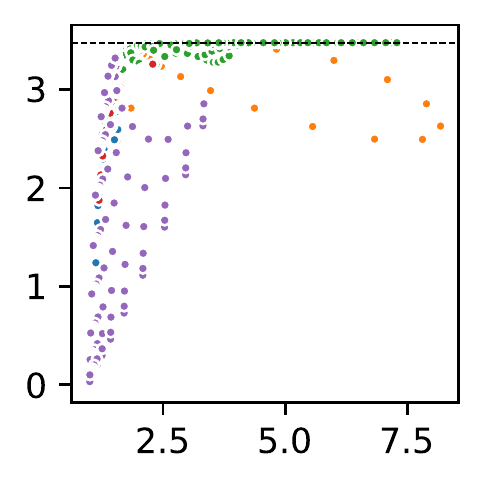}
  }
  \subfloat[Comp-Sweden]{
    \includegraphics[height=2.8cm]{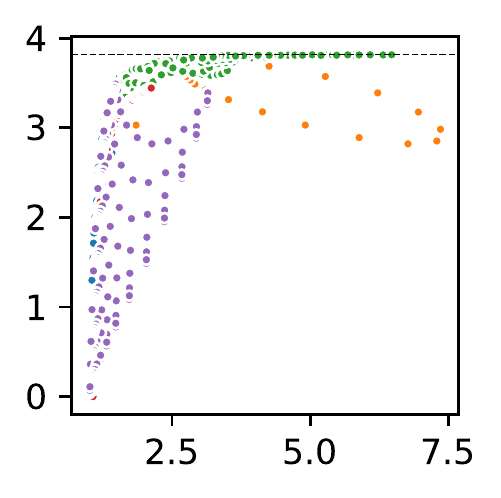}
  }      
  \subfloat[Comp-China]{
    \includegraphics[height=2.8cm]{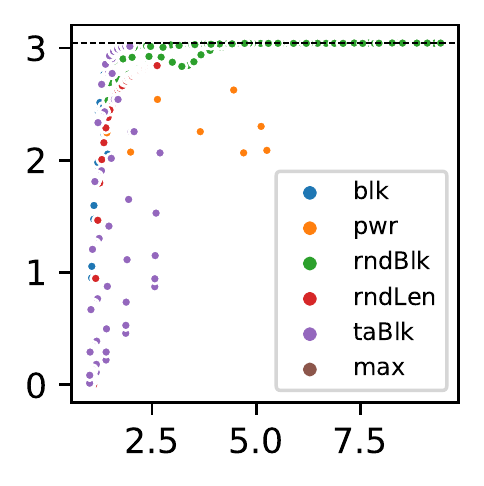}
  }    
  \subfloat[Comp-India]{
    \includegraphics[height=2.8cm]{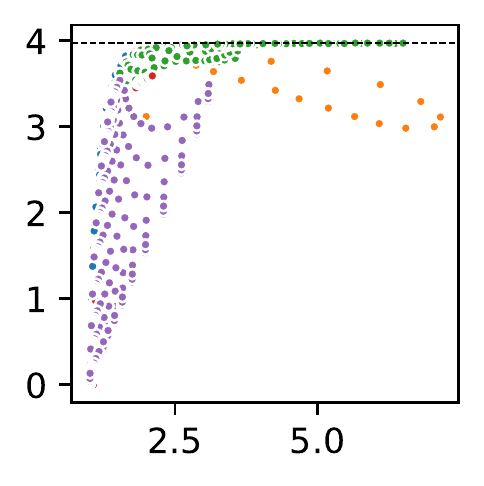}
  }    
  \subfloat[Comp-USA]{
    \includegraphics[height=2.8cm]{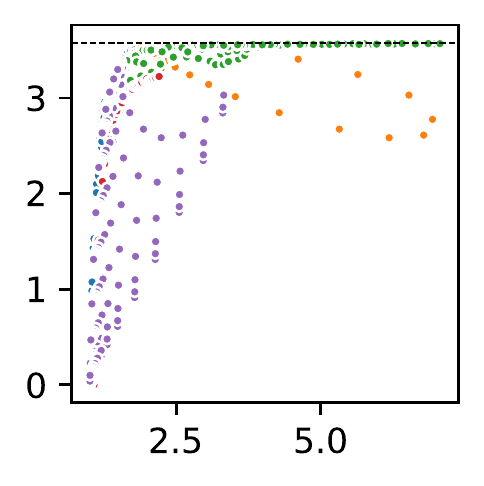}
  }  
  \caption{$\alpha_1$ vs $\beta$ plots. $\alpha_1$ in $y$-axis. $\beta$ in $x$-axis. Highest $\alpha_1=H(U)$ is indicated by the dashed horizontal line.\label{fig:defendervsattacker}}
  \end{figure*}

\begin{figure*}[h!]
  \centering  
  \subfloat[Sweden]{
    \includegraphics[height=2.8cm]{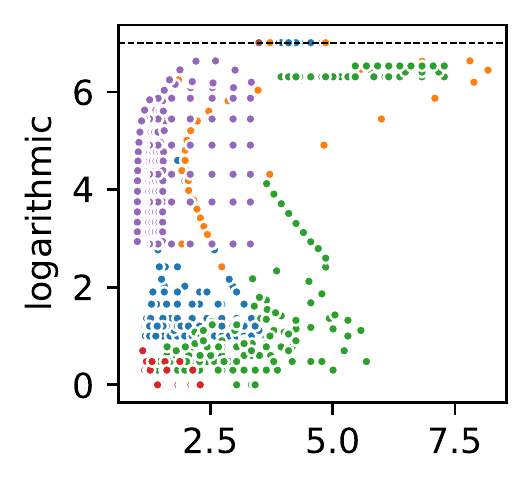}
  }
  \subfloat[Comp-Sweden]{
    \includegraphics[height=2.8cm]{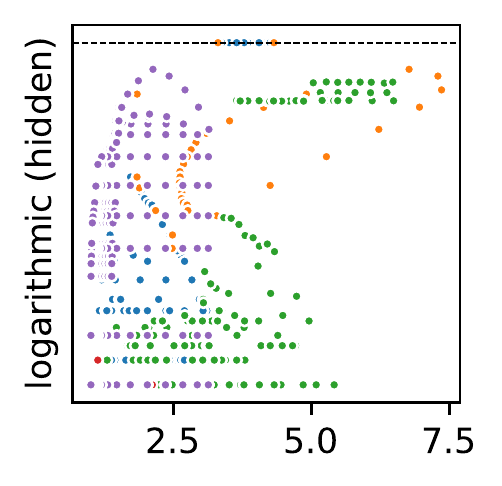}
  }    
  \subfloat[Comp-China]{
    \includegraphics[height=2.8cm]{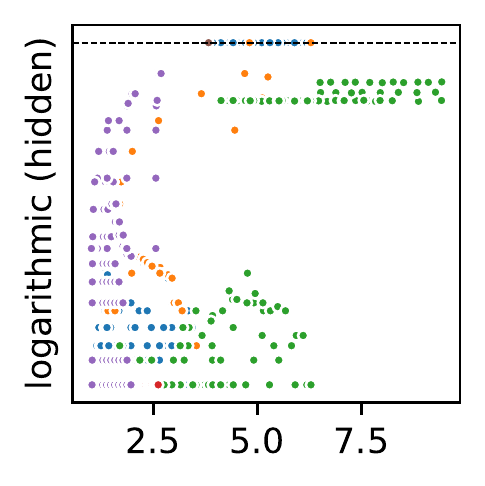}
  }  
  \subfloat[Comp-India]{
    \includegraphics[height=2.8cm]{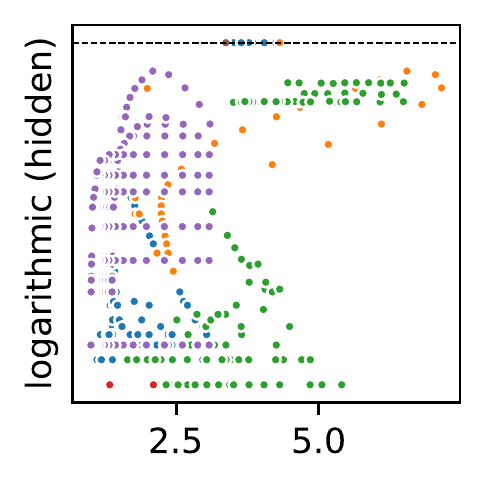}
  }  
  \subfloat[Comp-USA]{
    \includegraphics[height=2.8cm]{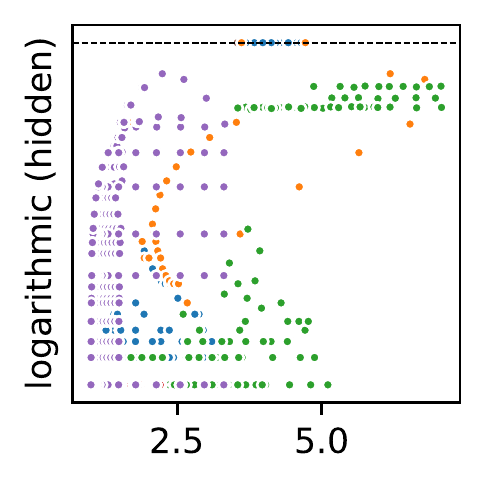}
  }    
  \caption{Log of $\alpha_2$ vs $\beta$ plots. Log of $\alpha_2$ in $y$-axis. $\beta$ in $x$-axis. Highest $\alpha_2=\texttt{population}$ is indicated by the dashed horizontal line.\label{fig:defendervsattacker2}}
\end{figure*}

\begin{figure*}[h]
  \centering
  \captionsetup[subfigure]{justification=centering}
  \subfloat[Sweden]{
      \includegraphics[height=3.7cm]{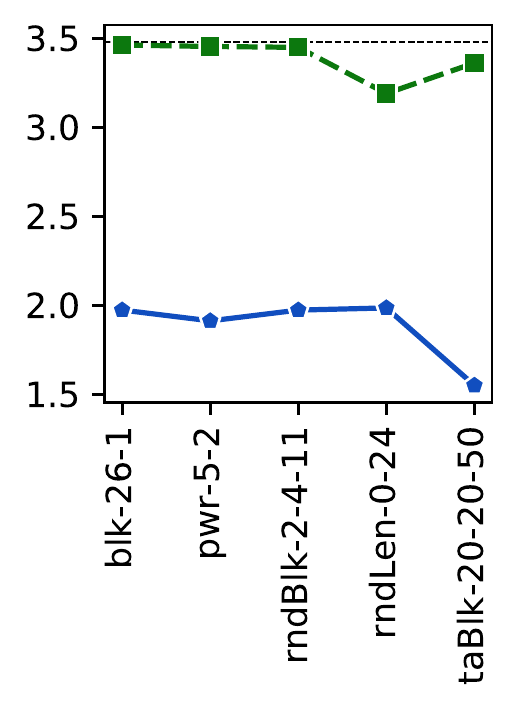}
  }
  \subfloat[Comp-Sweden]{
      \includegraphics[height=3.7cm]{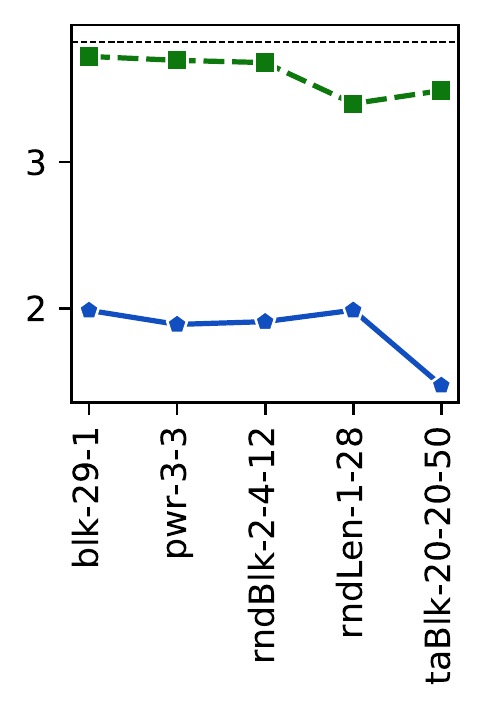}
  }
  \subfloat[Comp-China]{
      \includegraphics[height=3.7cm]{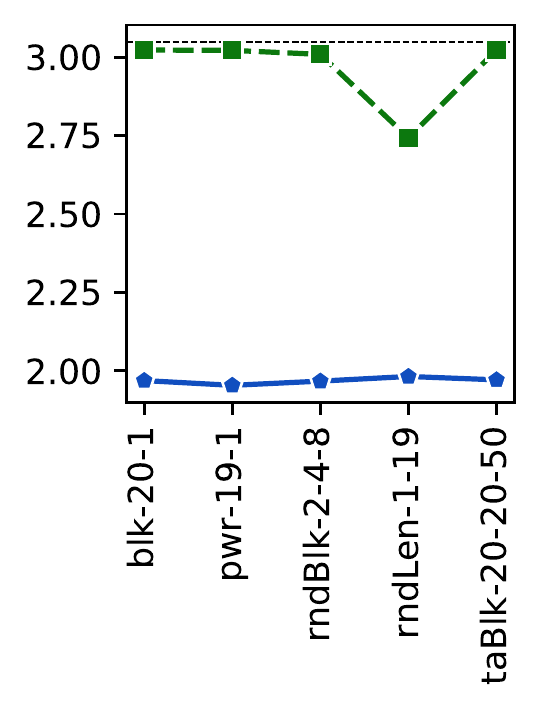}
  }
  \subfloat[Comp-India]{
      \includegraphics[height=3.7cm]{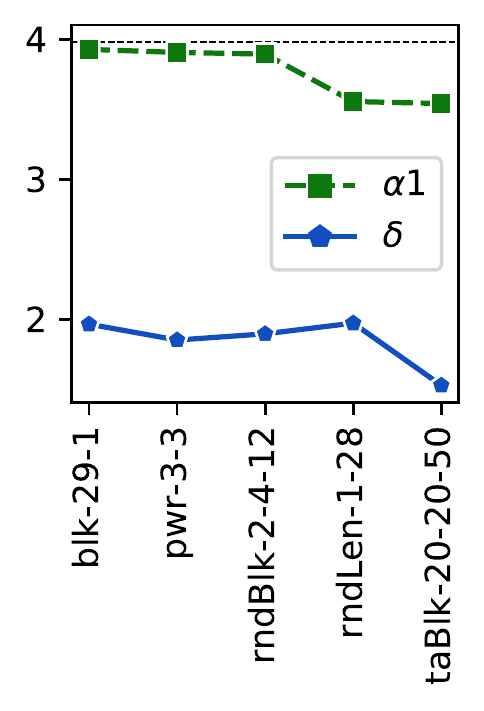}
  }
  \subfloat[Comp-USA]{
      \includegraphics[height=3.7cm]{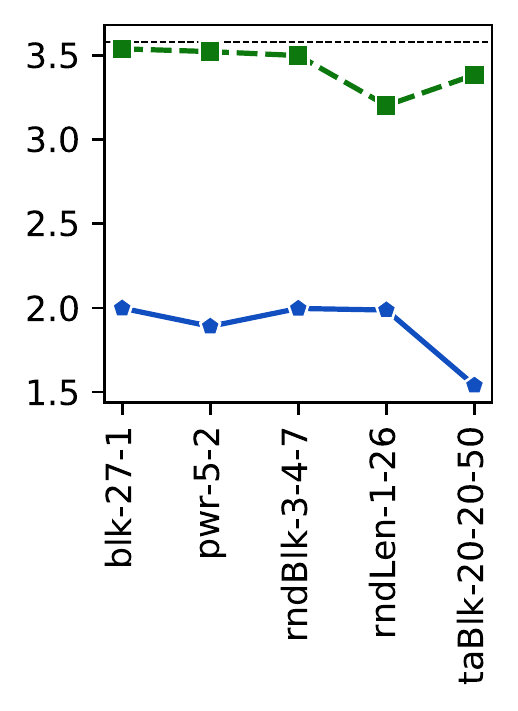}
  }  
  \caption{Best performing padding schemes in terms of lowest $\delta$ for $\beta\leq2$.\label{fig:compare_padding_schemes_defender}}
\end{figure*}

\begin{figure*}[h]
\centering
\captionsetup[subfigure]{justification=centering}
\subfloat[Sweden]{
    \includegraphics[height=3.7cm]{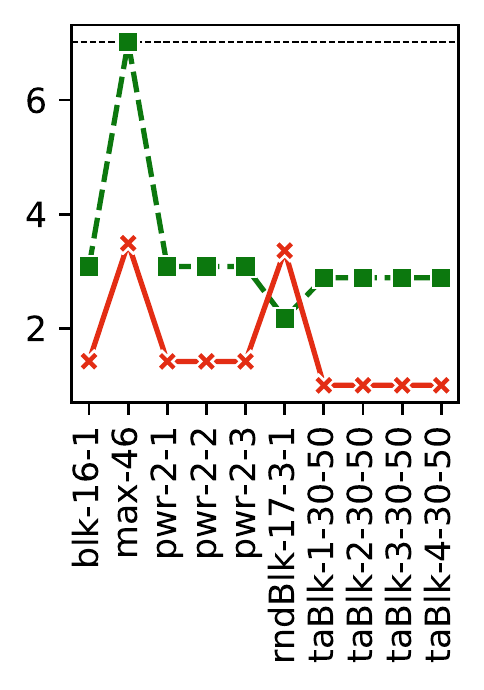}
}
\subfloat[Comp-Sweden]{
    \includegraphics[height=3.7cm]{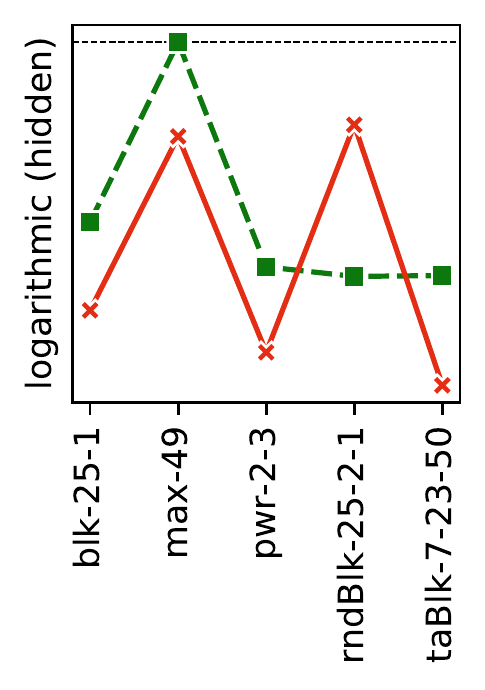}
}
\subfloat[Comp-China]{
    \includegraphics[height=3.7cm]{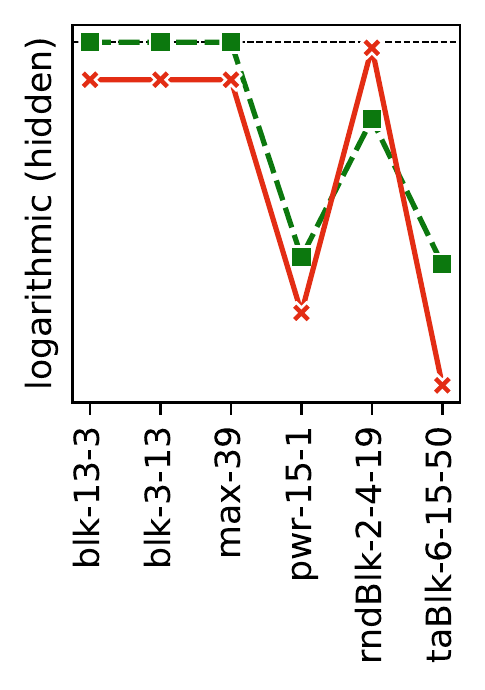}
}
\subfloat[Comp-India]{
    \includegraphics[height=3.7cm]{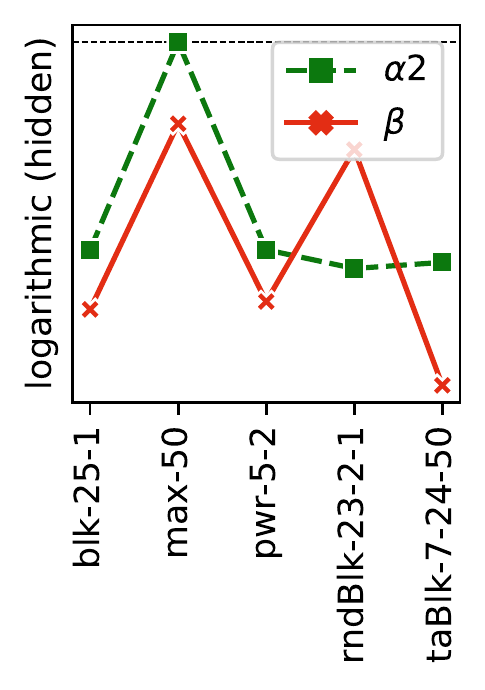}
}
\subfloat[Comp-USA]{
    \includegraphics[height=3.7cm]{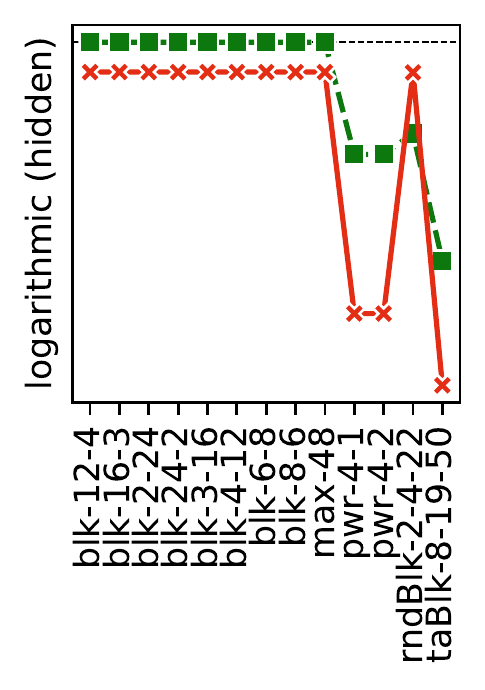}
}  
\caption{Best performing padding schemes in terms of lowest $\beta$ for $\alpha_2\geq100$.\label{fig:compare_padding_schemes_defender_2}}
\end{figure*}

\begin{itemize}[]
  \item \textbf{Distance to corner ($\delta$)}: For $\alpha_1$ vs $\beta$ plots in Fig. \ref{fig:defendervsattacker}, the points in the top-left corner are the instances where $\alpha_1$ is high and $\beta$ is low. Therefore, the points with low $\delta$ are better. Fig. \ref{fig:compare_padding_schemes_defender} shows the best performing padding schemes that have lowest $\delta$ for maximum $\beta$ of 2. It can be seen that $taBlk$ performs consistently better than others in terms of maintaining lower $\delta$. The distance metric used is the euclidean distance in the plotted 2-space, i.e., if the maximum $\alpha_1$ value is 4.0, the point $(1.0, 3.0)$ is assigned the distance $\delta = \sqrt{2}$. We note that the chosen distance metric is not necessarily the best; there could be other distance metrics giving different weights to $\alpha_1$ and $\beta$, and therefore giving different results. 
  
  \item \textbf{Threshold anonymity}: For $\alpha_2$ vs $\beta$ plots in Fig. \ref{fig:defendervsattacker2}, it could be sufficient to achieve some threshold anonymity $\alpha_2$. Fig. \ref{fig:compare_padding_schemes_defender_2} shows an example in which the threshold $\alpha_2$ is set to 100 and lowest $\beta$ required by padding schemes are identified. In this case too, $taBlk$ consistently performs better than others in terms of requiring lower $\beta$. We note that 100-anonymity may not provide acceptable anonymity in cases where an attacker has access to other out-of-band information.
\end{itemize}

\section{Recommendations\label{sec:reco}}
\textbf{Standardization.} We strongly recommend 3GPP and GSMA to mandate padding for all future SUCI generation profiles in 3GPP TS 33.501 \cite{3gpp_33501}. For such future schemes, we recommend including SUPI of both types (IMSI and NSI) for padding, because although IMSI is fixed length today, it may need expansion in future. We also recommend to introduce padding enabled versions of existing Profile A and B, because it is possible to do so in a backward compatible way. We further recommend that 3GPP and GSMA specify requirements and guidelines on how the impact of the Home network private key leakage can be mitigated, e.g., by limiting the number of subscribers that use a single public key (corresponding to the private key), and changing the private keys frequently.

\textbf{Padding Scheme}. We recommend the tail-aware block-length padding (\emph{taBlk-l-m-r}) scheme because of its superior performance in terms of $\alpha_1$, $\alpha_2$, $\beta$, and $\delta$. The choice of parameters ($l,m,r$) depends on name length distribution and how much message expansion ($\beta$) can be tolerated. \emph{taBlk-l-m-r} instances with $\beta \approx 2.0$ give good performance and excellent privacy. Furthermore, for Swedish name length data (Fig. \ref{fig:defendervsattacker2_zoom}), \emph{taBlk-l-m-r} instances with $\beta$ just above 1.0 (almost no average message expansion) gives 10000-anonymity which might be good enough for some deployments. It is hard to say what a minimum acceptable $k$-anonymity for 3GPP deployments are as an attacker might use additional out-of-band information like realm, location, time, and cell tower identifier in identifying subscribers. A realm that is used over a large geographical area should have a much higher $k$-anonymity than a realm used in a small geographical area. As it is difficult to analyze all the out-of-band information that an attacker might have access to, we recommend that the targeted level of $k$-anonymity is chosen with large margins. For roaming users and small companies with their own realm, good anonymity is likely not possible even with padding. If non-ASCII characters are used in NAIs without transliteration to ASCII, the distributions given in this paper might change substantially. In particular such distribution could in the worst case have low (non-zero) frequencies in the middle of the distribution. A slightly modified tail-aware block-length padding with a bit block size larger ($>1$) in the middle would likely be an optimal choice also for those distributions. 

According to the type of radio access technology, the messages carrying SUCI have maximum size from 1600 to 9000 bytes \cite{3gpp_24501, 3gpp_38323, 3gpp_36323}. In all our datasets, the maximum username length we observed is 50. Given this and other sources (e.g., maximum 70 in e-GIF \cite{uk_length} and 64 in SMTP \cite{rfc5321}), accepting $\beta \approx 2.0$ seems to be safe and still allow plenty of space for padding. 
\section{Ethics and Responsible Disclosure}
The data from Sweden and Stockholm were already anonymized by SCB; the authors only received a frequency table stating how many people have a name with a certain length. Similarly, the data from the multi-national company were anonymized to contain only the lengths of the username part of the email addresses. To anonymize the Company's data further, the name of the Company is not revealed in this paper, and corresponding plots are either normalized or shown in a logarithmic scale. 

We have done responsible disclosure of our vulnerability discovery and solution proposal to the 5G Security Task Force of GSMA’s Fraud and Security Group (FASG) \cite{gsma_security}. GSMA FASG acknowledged that the vulnerability is real and should be fixed. We received positive gestures that this work may be pursued in 3GPP, aiming for Release-18.
\section{Related Work}
Real-world data sets of name lengths are surprisingly hard to find in the open literature.  Only a few earlier works \cite{Healy1968TheLO, Jackson2013FittingAD} have investigated the statistical distribution of the length of names. 

IPsec ESP packets \cite{rfc4303} and TLS 1.3 records \cite{rfc8446} provide mechanisms to add padding to the plaintext before encryption. They however do not provide any padding policies, leaving that to the implementation. TLS Encrypted Client Hello \cite{ietf-tls-esni-09} recommends a padding scheme where small identities are padded to a fixed length and long identities are padded to the nearest multiple of 32 bytes. Padding for DNS request and response messages are specified in \cite{rfc7830} based on empirical research and recommendations in \cite{ndss, rfc8467}. In \cite{Karopoulos2010AFF}, the authors discussed using padding schemes -- OAEP for RSA and ISO 10126 for AES. The authors did not suggest any padding for ECIES since the encryption key itself is different every time, therefore, the purpose of padding in \cite{Karopoulos2010AFF} is to ensure the so-called probabilistic encryption and does not cater privacy leakage through length of encrypted identifier.

In \cite{Liu2014PPTPPT}, the authors propose a formal method for quantifying the amount of privacy protection provided by traffic padding solutions. The model encompasses the privacy requirements, padding costs, and padding methods. This enables applications to choose more optimal padding methods, minimizing padding costs for a given amount of privacy, or maximizing the amount of privacy for a given padding cost. Wagner et al. \cite{privacy_metric} give an overview of privacy metrics that have
been proposed in the literature and classify them based on the aspect of privacy they measure. They conclude that the lack of standardized privacy metrics makes it hard makes the choice of privacy metric and comparison between different studies hard. 

Mathur et al. \cite{5953511} analyse the privacy protection given by padding and encryption of traffic data sent over a network. They conclude that previous work has been too focused on padding to a fix length, which waste unnecessary amounts of bandwidth. They recommend a randomized padding and use conditional entropy as a privacy metric to evaluate the protection. 
\section{Conclusion and Further Work}
5G SUCI uses state-of-the-art cryptography and gives very good privacy protection when used on fixed length identifiers such as IMSI. But when used on variable length identifiers such as NSI, it leaks significant amount of information that may be practically useful for an attacker to identify and track a victim. We presented real-world name length data, discussed padding as a solution, and showed empirical evaluation of several padding schemes. We have presented our work to GSMA FSAG's 5G Security Task Force \cite{gsma_security}. We strongly recommend 3GPP and GSMA to pursue this work and standardize a padding mechanism for all SUPI types in 5G.

Future research should consider the name distributions in which usernames are not transliterated to ASCII characters; such distributions could in the worst case have low frequencies in the middle of the distribution and require a slightly different padding scheme. It might also prove important to measure the message sizes that the identifiers are transported in; if these messages are large, the bandwidth overhead as measured in percentage of the original message could be small and more padding might be acceptable. Future investigations are also necessary on how large typical 5G realms will be, how much additional information (e.g., realm, location, time, and other data) attackers can use, and therefore what acceptable minimum values for $k$-anonymity in various deployments should be.

\section*{Acknowledgements}
  We thank Ann-Marie Persson for enabling our research with SCB data and Daniel Kahn Gillmor for providing technical comments. We also thank our Ericsson colleagues for reviewing the manuscript and especially Erik Thormarker for brainstorming ideas with us.

\bibliographystyle{acm}
\bibliography{./tex_arc/88.references}

\appendix
\newpage
\section{SCB Data} \label{sec:scb_data}
\begin{table}[h]
	\footnotesize
	\centering
    \caption{Frequency tables for name lengths in Sweden and Stockholm -- as of 31 December 2019 -- in the forms first name $||$ last name (fl) and first name $||$ maiden name $||$ last name (fml). Source: Swedish government agency SCB (Statistiska centralbyrån) \cite{SCB}.}
	\label{tab:scb_data}
	\centering
	\csvreader[
		tabular=|r|r|r|r|r|,
		table head=\hline \bfseries{Length} & \bfseries{Swe-fl} & \bfseries{Swe-fml} & \bfseries{Sthlm-fl} & \bfseries{Sthlm-fml} \\\hline,
		late after last line=\\\hline 
	  ]{
		"anc/scb_name_length_data_Sweden_Stockholm_2019.csv"
	  }{}{\csvlinetotablerow}
\end{table}

\end{document}